\documentclass[aps,prr,twocolumn,
groupedaddress]{revtex4-1}
\usepackage{amsmath,amssymb,graphicx,stmaryrd}
\usepackage{physics}
\usepackage[utf8]{inputenc}
\usepackage[T1]{fontenc}
\usepackage{xcolor}
\usepackage{comment}

\IfFileExists{newtxtext.sty}
   {\usepackage{newtxtext,newtxmath}}
   {\IfFileExists{stix.sty}
      {\usepackage{stix}}
      {\IfFileExists{mathptmx.sty}
      {\usepackage{mathptmx}}{} } }

\usepackage{booktabs,siunitx}

\pdfoutput=1
\definecolor{LinkColor}{rgb}{0.256,0.0439,0.588}
\usepackage{hyperref}
\hypersetup{
   pdfauthor={Sudeep Adhikari and K. S. D. Beach},
   pdftitle={Universal aspects of barrier crossing under bias},
   pdfsubject={Escape rate from a confining well},
   pdfkeywords={single-molecule pulling experiments,} {optical tweezers,} {energy landscape,} {escape rate,} {rupture force distribution,} {Langevin dynamics,} {diffusion},
   colorlinks=true,
   citecolor=LinkColor,
   linkcolor=LinkColor,
   urlcolor=LinkColor
   }

\newcommand{\linkdoi}[2]{\href{http://dx.doi.org/#2}{#1}}
\newcommand{\linkisbn}[2]{\href{https://isbndb.com/book/#2}{#1}}

\begin{document}

\title{Universal aspects of barrier crossing under bias}

\author{Sudeep Adhikari}
\email[Electronic mail:\ ]{sadhika6@go.olemiss.edu}
\affiliation{Department of Physics and Astronomy, The University of Mississippi, University, Mississippi 38677, USA}

\author{K. S. D. Beach}
\email[Electronic address:\ ]{kbeach@olemiss.edu}
\affiliation{Department of Physics and Astronomy, The University of Mississippi, University, Mississippi 38677, USA}

\date{March 30, 2022}

\begin{abstract}
The thermal activation process by which a system passes from one local 
energy minimum to another is a recurring motif in physics, chemistry, 
and biology. For instance, biopolymer chains are typically modeled in 
terms of energy landscapes, with folded and unfolded conformations 
represented by two distinct wells separated by a barrier. The rate of 
transfer between wells depends primarily on the height of the barrier, 
but it also depends on the details of the shape of the landscape along 
the trajectory. We consider the case of bias due to an external force,
analogous to the pulling force applied in optical tweezer experiments on
biopolymers. Away from the Arrhenius-law limit and well out of equilibrium, 
somewhat idiosyncratic behavior might be expected. Instead, we identify
universal behavior of the biased activated-barrier-crossing process and
demonstrate that data collapse onto a universal curve can be achieved for
simulated data over a wide variety of energy landscapes having barriers
of different height and shape and for loading rates spanning many orders
of magnitude.
\end{abstract}
\maketitle

\section{Introduction \label{SEC:introduction}}

Thermally activated barrier crossing~\cite{Kramers-Phy-40, Grote-JCP-80, Pollak-JCP-86, Hanggi-RMP-90} is a ubiquitous 
and highly consequential process in physics, chemistry, and biology. An understanding of 
the factors that influence the rate of barrier crossing~\cite{Fleming-JSP-86,Haynes-JCP-94,Kraikivski-EL-04,Makarov-JPC-21}
is necessary for the interpretation of experiments that attempt to infer barrier height and 
shape from measurements of the escape rate. An important specialization is the 
case in which the barrier is diminished by an applied force, with the escape rate
enhanced accordingly.

Experimental access to escape rate information in the biochemistry context has been revolutionized by
the development of single-molecule force 
spectroscopy~\cite{Rief-Science-97,Kellermayer-Science-97,Tskhovrebova-Nature-97,Bustamante-ARBC-04,Greenleaf-ARBBS-07,Hyeon-JPCM-07,Bustamante-ARBC-08,Neuman-NatMeth-08,Woodside-COCB-08,Zoldak-COSB-13}, 
in which a mechanical load is applied across a single molecule using an atomic force 
microscope or optical tweezers. In the energy landscape picture~\cite{Bryngelson-PNAS-87, Onuchic-AnnRevPhysChem-97, Onuchic-AdvProtChem-00, Wales-Science-01, Onuchic-COSB-04, Lacks-BPJ-05},
molecular motion is viewed as a Brownian diffusion process over a
free energy surface~\cite{Dill-Science-12}, parameterized by the conformational 
degrees of freedom. 
The landscapes for biologically relevant sequences contain distinct, barrier-separated wells corresponding to various folded and unfolded conformations. 
The rate of transition~\cite{Szabo-JCP-80,Chung-PNAS-09,Chaudhury-JCP-10} from one well to another depends primarily 
on the height of the intervening barrier but also depends
on its shape.

In pulling experiments, where molecular extension serves as a natural
reaction coordinate, the multidimensional energy landscape
covering the full comformational space is projected onto an
effective one-dimensional energy profile that encodes some features
of the full landscape and that reproduces the folding dynamics~\cite{Kirmizialtin-JCP-05, Woodside-PNA-06, Best-PNAS-10, Yu-PNAS-12, Patten-CPC-17}.
Numerous studies have been carried out
to explore the unfolding process under the application of constant and time-varying pulling 
forces~\cite{Bell-Science-78,Evans-BPJ-91,Li-JCP-04,Souza-NatMethods-12,Ritchie-COSB-15,Woodside-BPJ-14,Neupane-NAR-11,Bull-ACSNano-14,Rief-PRL-98,Hummer-BPJ-03,Friddle-PRL-08}. 
A key experimental goal
is to be able to reliably reconstruct the effective one-dimensional free energy profile from measurements
of an ensemble of escape events~\cite{Hummer-PNAS-01,Gebhardt-PNAS-10,Gupta-Nature-11,Messieres-BJ-11,Woodside-ARB-14,Engel-PRL-14}.

There are additional complications that may arise
because of the multidimensional nature of the landscape~\cite{Dudko-PNAS-03,Yew-PRE-10,Suzuki-PRL-10,Dudko-PRL-11,
Avdoshenko-JPC-15,Makarov-JCP-16}, but
we consider the common case in which the landscape can be meaningfully
projected onto a one-dimensional effective energy in a well-chosen reaction coordinate~\cite{Dudko-PRL-11}.

The purpose of this article is to describe universal aspects of the biased activated-barrier-crossing process that we have 
uncovered in numerical simulations of various one-dimensional potentials. Our work points the way to an alternative data analysis 
technique that would allow for the determination of otherwise unknown landscape details
by overlaying data from multiple experiments 
and adjusting free parameters until the scattered data align
along a common curve.

The concept of universality comes to us from the study of
critical phenomena~\cite{Stanley-RMP-99}. In that context it allows us to understand how
phase transitions can be characterized and grouped into families according to common critical exponents, wholly independent of the microscopic details of the underlying models; it also explains the existence of scaling relations that govern
how thermodynamic quantities behave in the vicinity of criticality.
An important mark of universality is that data from different models or different physical systems can be plotted in reduced variables so that they collapse onto a single universal curve~\cite{Kawashima-JPSJ-93,Bhattacharjee-JPA-01,Houdayer-PRB-04}.

Criticality has previously been invoked by Singh, Krishan, and Robinson
in the context of the unbiased-activated-barrier crossing problem~\cite{Singh-PRL-92,Singh-PRE-94}.
They considered the non-Markovian crossing of a quadratic barrier,
where the frictional term in the Langevin equation includes a memory kernel with a
long time scale. The authors proposed a scaling hypothesis, making
analogy with the criticality of the Ising model,
and were able to derive scaling relations for the reduced rate
near a critical value of the memory kernel time scale.

Our approach here is rather different.
We focus on the relative change in the escape rate
as a function of an applied pulling force---both for uniform
pulling ($F$ constant) and steady loading ($F=KVt$ with $KV$ constant).
We propose that $F$ exists alongside two other important force scales
and that the two independent ratios that can be formed serve as arguments to a scaling function. 
We have carried out Langevin-type simulations of a particle in a one-dimension
energy potential, coupled to a heat bath. Many thousands of instantiations
provide us with a large data set that offers good coverage of the model space.
What is so striking is the almost unreasonable effectiveness of the scaling ansatz, 
which appears to be valid over a huge variety of well shapes and barrier heights and over
loading rates spanning many orders of magnitude.

\begin{figure}
\begin{center}
\includegraphics{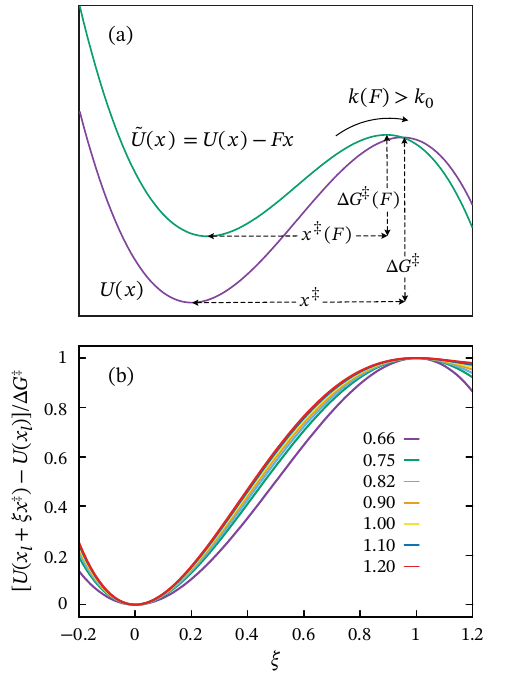}
\end{center}
\caption{\label{FIG:Double-well} 
(a) In the unbiased energy landscape [solid purple line, labelled $U(x)$], 
a particle escaping (left to right) from the well must traverse a barrier 
of height of  ${\Delta G}^\ddagger = U(x_b) - U(x_l)$, over a distance 
$x^\ddagger = x_b - x_l$, where $x_l$ is the position of the bottom of the 
left well and $x_b$ is the position of the barrier peak. With application of 
an assistive pulling force ($F>0$), the energy landscape tilts 
(solid green line) to favor the destination well to the right of the barrier.
The pulling force causes the extrema to shift; the barrier height 
${\Delta G}^\ddagger(F)$ and barrier distance $x^\ddagger(F)$ both decrease. 
(b) Seven energy profiles are depicted, rescaled so that the bottom of the 
well and top of the barrier coincide. The color key shows the shape parameter 
($\nu$) values for each curve. }
\end{figure}

\section{Scaling Ansatz \label{SEC:scaling-ansatz}}

We argue that the barrier-crossing process is controlled by the relative magnitudes of three intrinsic force scales: 
the typical thermal force that provides the kick out of the well ($F_T \approx 1/\beta x^\ddagger$);
the larger applied force required to fully extinguish the barrier ($F_B \approx \kappa^\ddagger x^\ddagger$);
and the pulling force used as an external bias $(F)$. In our notation, $\beta$ is the inverse temperature, and
$x^\ddagger$, $\kappa^\ddagger$ are the barrier distance (see Fig.~\ref{FIG:Double-well}) and effective curvature~\cite{Adhikari-PRR-20}.
In particular, we propose that the rescaled, logarithmic relative escape rate $(F_T/F_B)\ln [k(F)/k_0]$, when plotted against the reduced pulling force $F/F_B$, collapses onto a universal curve. 
The function $k(F)$ is the escape rate associated with the potential landscape under bias, 
and $k_0$ is the corresponding rate in the untilted landscape.

More precisely, the claim is that  
\begin{equation} \label{EQ:scaling-form}
\begin{split}
Y(F) &= \ln \frac{k(F)}{k_0} = \biggl(\frac{F_T}{F_B}\biggr)^{-1}
\mathbb{Y}(F/F_B, F_T/F_B)\\
&= \biggl(\frac{F_T}{F_B}\biggr)^{-1}
\biggl[\mathbb{Y}(F/F_B, 0) \\
&\quad\qquad\qquad + \biggl(\frac{F_T}{F_B}\biggr)^\omega \mathbb{W}(F/F_B) + \cdots \biggr].
\end{split}
\end{equation}
Up to subleading corrections (characterized by an exponent $\omega > 0$),
the escape behaviour is controlled by a universal function $\mathbb{Y}(x, y)$
that satisfies $\mathbb{Y}(x, 0) \sim x + O(x^2)$. 
The implication of Eq.~\eqref{EQ:scaling-form} is that
a plot of $(F_T/F_B)Y$ versus $f = F/F_B$ should produce 
data collapse regardless of the microscopic details of the simulation.

In the context of dynamical pulling, there is another useful analysis.
A population of particles trapped in the originating well is depleted 
according to $-dn/dt = k(KVt)n(t)$, where the right-hand side of the
equality is a product of the instantaneous escape rate and the current population. The population's half-life is characterized by
$\ln 2 = \int_1^{1/2}\!dn/n = - (1/KV)\int_0^{\hat{F}}\!dF\,k(F)$. Here,
$\hat{F} = KV\hat{t}$ is the typical applied force that is
in effect during barrier transit, and $\hat{t}$ is the median elapsed time for escape.
It follows from Eq.~\eqref{EQ:scaling-form} that $\hat{F}$ (measured with respect to the thermal force $F_T$)
must be a monotonic, universal function of 
$\dot{F} = KV$ (measured with respect to $k_0 F_T$, a loading rate threshold defined 
by the thermal processes in the potential well). Hence, there
is an additional data collapse analysis that can be used to independently
test the validity of the scaling hypothesis.

In our numerical experiments, the external bias is applied in two ways:
(i) as a time-invariant pulling force of constant strength and (ii) as a linearly time-varying force 
with a constant loading rate.
In the case of constant pulling, the system is prepared in the equilibrium state of the tilted energy profile
[viz.\ $\tilde{U}(x) = U(x) - Fx$] and remains in thermal equilibrium throughout the simulation. In the case of steady loading, the system is prepared in the equilibrium state of the unbiased profile ($F=0$ for all $t\le 0$), but as time elapses it is driven away from equilibrium ($F = KVt$ for all $t>0$) in proportion to how much $KV$ exceeds $k_0F_T$.

In both cases, the role of $F$ is to gently tilt the landscape (statically 
or dynamically), as depicted in the upper panel of Fig~\ref{FIG:Double-well}.  
There, the purple curve depicts the potential profile in its unbiased state;
the green curve shows the profile after application of the external bias. 
Generically, the force-dependent values of the barrier distance $x^\ddagger(F)$
and barrier height ${\Delta G}^\ddagger (F)$ are monotonic decreasing in $F$, 
and hence the barrier crossing process becomes energetically less costly (and 
crossing events more frequent) as the external bias is ramped up.
The free-energy minimum is progressively destabilized
and disappears entirely at the threshold for barrier extinction.

\section{Theoretical motivation \label{SEC:theoretical-motivation}}

\subsection{Locally quadratic approximation}\label{SUBSEC:locally-quadratic-approx}

We consider a one-dimensional, double-well energy landscape $U(x)$ with minima
on the left and right, at positions $x_l$ and $x_r$, separated
by a barrier at $x_b$. A barrier of height $\Delta G^\ddagger = U(x_b) - U(x_l)$
impedes transitions from left to right.
By definition $U'(x_l) = U'(x_b) = U'(x_r) = 0$.
In the locally quadratic approximation, we assume
\begin{equation} \label{EQ:quadratic-approximation}
U(x) = \begin{cases}
U(x_l) + \frac{1}{2}\kappa_l(x-x_l)^2 & \text{for $x \simeq x_l$}, \\
U(x_b) - \frac{1}{2}\kappa_b(x-x_b)^2 & \text{for $x \simeq x_b$},
\end{cases}
\end{equation}
where $\kappa_l = U''(x_l)$ and $\kappa_b = -U''(x_b)$
are measures of the curvature at the bottom of the well
and at the top of the barrier.

With the application of a bias force $F$, the extrema of the tilted landscape 
$\tilde{U}(x) = U(x) - Fx$ are found as follows:
\begin{equation}
0 = \tilde{U}'(x) = U'(x) - F 
= \begin{cases}
+ \kappa_l(x-x_l)-F,\\
- \kappa_b(x-x_b)-F.
\end{cases}
\end{equation}
At this level of approximation, the bias-induced shifts in the extrema are linear in $F$.
In response to the applied force ($F>0$), the well basin moves to the right and the barrier peak
moves to the left:
\begin{equation}
\tilde{x}_l = x_l + \frac{F}{\kappa_l},~~~\tilde{x}_b = x_b - \frac{F}{\kappa_b}.
\end{equation}
The inverse spring constants, $1/\kappa_l$
and $1/\kappa_b$, represent the compliance of the reactant and
transition state; see Eq.~(4) of Ref.~\onlinecite{Makarov-JCP-16}
and the accompanying discussion. The two points eventually coalesce 
when $\tilde{x}_l = \tilde{x}_b$, i.e., when
\begin{equation} \label{EQ:quadratic-barrier-distance}
x^\ddagger = x_b - x_l = \biggl(\frac{1}{\kappa_l} + \frac{1}{\kappa_b}\biggr)F
\equiv \frac{F}{\kappa^\ddagger}.
\end{equation}
The particular force value at which Eq.~\eqref{EQ:quadratic-barrier-distance} 
holds is the barrier extinction force $\kappa^\ddagger x^\ddagger$. 
We follow the usual practice of decorating with a double-dagger superscript 
any quantity that is defined with respect to the barrier and the originating well.
This includes the barrier distance $x^\ddagger = x_b - x_l$
and the effective curvature
\begin{equation}
\kappa^\ddagger = \biggl(\frac{1}{\kappa_b} + \frac{1}{\kappa_l}\biggr)^{-1} 
\!= \frac{\kappa_l\kappa_b}{\kappa_l+\kappa_b}.
\end{equation}

In order to find an expression for the barrier height that is
consistent with the approximation in Eq.~\eqref{EQ:quadratic-approximation}, 
we must match the two piecewise quadratic curves. We do
so at the point of common slope, where
\begin{equation}
U'(x^*) = \kappa_l(x^*-x_l) = -\kappa_b(x^*-x_b).
\end{equation}
The reference position
\begin{equation} \label{EQ:common-slope}
x^* = \frac{\kappa_l x_l + \kappa_b x_b}{\kappa_l + \kappa_b}
\end{equation}
is a weighted average satisfying $x_l \le x^* \le x_b$.
The height of the barrier in the untilted landscape ($F=0$) is estimated to be
\begin{equation} \label{EQ:barrier-height-quadratic}
\begin{split}
\Delta G^\ddagger
&= U(x_b) - U(x^*) + U(x^*) - U(x_l)\\
&\approx \frac{1}{2}\kappa_b(x^*-x_b)^2 + \frac{1}{2}\kappa_l(x^*-x_l)^2
= \frac{1}{2}\kappa^\ddagger {x^\ddagger}^2.
\end{split}
\end{equation}
Formally, the barrier extinction force is given
by the derivative of the barrier height with respect to the barrier position.
At the level of approximation of Eq.~\eqref{EQ:barrier-height-quadratic},
we have
\begin{equation} \label{EQ:extinction-force-quadratic}
\frac{\partial \Delta G^\ddagger}{\partial x^\ddagger} = \kappa^\ddagger x^\ddagger.
\end{equation}

\subsection{Higher-order corrections}
\label{SUBSECT:higher-order-corrections}

There are various forms of ``extended Bell theory''~\cite{Huang-PAC-10,Konda-JCP-11,Bailey-JCP-12} that offer systematic refinements to the transition rate under bias. These are generally structured order by order in the applied force, as per Eq.~(1) of Ref.~\onlinecite{Huang-PAC-10}.
Indeed, we can improve on the derivation in Sec.~\ref{SUBSEC:locally-quadratic-approx} by including further contributions to the energy landscape expansion:
\begin{equation} \label{EQ:higher-order-corrections}
U(x) = \begin{cases}
U(x_l) + \frac{1}{2!}\kappa_l(x-x_l)^2 - \frac{1}{3!}R_l(x-x_l)^3 \\ \qquad\qquad\qquad{}+ \frac{1}{4!}Q_l(x-x_l)^4 - \cdots,\\
U(x_b) - \frac{1}{2!}\kappa_b(x-x_b)^2 - \frac{1}{3!}R_b(x-x_b)^3 \\ \qquad\qquad\qquad{}- \frac{1}{4!}Q_b(x-x_b)^4 - \cdots.
\end{cases}
\end{equation}
As in Eq.~\eqref{EQ:quadratic-approximation}, the upper expression
in Eq.~\eqref{EQ:quadratic-barrier-distance}
is for $x \simeq x_l$; the lower corresponds to $x \simeq x_b$.
In addition to the two local curvatures, $\kappa_l$ and $\kappa_b$, 
we have also defined measures of the skew [$R_l = -U'''(x_l) = -U^{(3)}(x_l)$ and $R_b = -U^{(3)}(x_b)$]
and the kurtosis [$Q_l = U^{(4)}(x_l)$ and $Q_b = -U^{(4)}(x_b)$].

The positions of the shifted extrema are once again determined by $0 = U'(x)-F$.
This demands that the expression
\begin{multline}
~~~ s\kappa_\alpha(x-x_\alpha) - \frac{1}{2}R_\alpha(x-x_\alpha)^2\\ + \frac{s}{6}Q_\alpha(x-x_\alpha)^3 + \cdots - F ~~~
\end{multline}
vanish for both $\alpha = l$, $s=+1$ and $\alpha = b$, $s = -1$.
Ensuring that it does so leads to
\begin{equation}
\tilde{x}_\alpha = x_\alpha + \frac{sF}{\kappa_\alpha}
+ \frac{sR_\alpha F^2}{2\kappa_\alpha^3}
+ \frac{s(3 R_\alpha^2 - Q_\alpha\kappa_\alpha)F^3}{6\kappa_\alpha^5} + \cdots
\end{equation}
and hence to an expression for $\tilde{x}_b - \tilde{x}_l$, the barrier distance
in the tilted energy landscape:
\begin{equation}
x^\ddagger - \frac{F}{\kappa^\ddagger}
- \frac{R^\ddagger_{1,3} F^2}{2{\kappa^\ddagger}^3}
- \biggl(\frac{3\bigl(R^\ddagger_{2,5}\bigr)^2}{{\kappa^\ddagger}^5} 
- \frac{Q^\ddagger_{1,4}}{{\kappa^\ddagger}^4}\biggr)\frac{F^3}{6} + \cdots.
\end{equation}
As a convenience, we have adopted the notation
\begin{equation}
A^\ddagger_{m,n} = 
{\kappa^\ddagger}^n\bigg(\frac{A_l^m}{\kappa_l^n} + \frac{A_b^m}{\kappa_b^n}\biggr) =
\frac{A_l^m\kappa_b^n + A_b^m\kappa_l^n}{(\kappa_l+\kappa_b)^n}.
\end{equation}

If we then repeat the analysis used previously, we can produce expressions
for the barrier height and barrier extinction force that are
analogs
of Eqs.~\eqref{EQ:barrier-height-quadratic} and \eqref{EQ:extinction-force-quadratic}:
\begin{multline} \label{EQ:barrier-height}
\Delta G^\ddagger = \frac{1}{2!}\kappa^\ddagger {x^\ddagger}^2
- \frac{1}{3!}R^\ddagger_{1,3}{x^\ddagger}^3 \\
+ \frac{1}{4!}\biggl(\frac{3}{\kappa^\ddagger}\bigl[(R^\ddagger_{1,3})^2-R^\ddagger_{2,5}\bigr]
+ Q^\ddagger_{1,4}\biggr){x^\ddagger}^4 + \cdots
\end{multline}
and
\begin{multline} \label{EQ:extinction-force}
\frac{\partial \Delta G^\ddagger}{\partial x^\ddagger} = \kappa^\ddagger x^\ddagger
- \frac{1}{2}R^\ddagger_{1,3}{x^\ddagger}^2 \\
+ \biggl(\frac{(R^\ddagger_{1,3})^2}{2\kappa^\ddagger}
- \frac{R^\ddagger_{2,5}}{2\kappa^\ddagger} 
+ \frac{1}{6}Q^\ddagger_{1,4}\biggr){x^\ddagger}^3 + \cdots
\end{multline}

It is helpful to distinguish the barrier height expressions in 
Eqs.~\eqref{EQ:barrier-height-quadratic} and \eqref{EQ:barrier-height}
by the labels $\Delta G^\ddagger_{\!\text{quad}}$ and $\Delta G^\ddagger$.
Their ratio is simply the {\it shape parameter} defined
by Dudko, Hummer, and Szabo~\cite{Dudko-PRL-06}:
\begin{equation} \label{EQ:shape-parameter}
\begin{split}
\nu = \frac{\Delta G^\ddagger}{ \Delta G^\ddagger_\text{quad}} 
&= \frac{\frac{1}{2}\kappa^\ddagger {x^\ddagger}^2 - \frac{1}{6}R^\ddagger_{1,3} {x^\ddagger}^3 + \cdots }{\frac{1}{2}\kappa^\ddagger {x^\ddagger}^2}\\
&= 1 - \frac{R^\ddagger_{1,3} x^\ddagger}{3\kappa^\ddagger} + 
\biggl(\frac{(R^\ddagger_{1,3})^2}{4{\kappa^\ddagger}^2}\\
&\qquad\qquad - \frac{R^\ddagger_{2,5}}{4{\kappa^\ddagger}^2} 
+ \frac{Q^\ddagger_{1,4}}{12\kappa^\ddagger}\biggr){x^\ddagger}^2 +
\cdots 
\end{split}
\end{equation}
That is to say, $1-\nu$ encodes deviations from the behaviour of the purely quadratic model
(in which $R_l = R_b = 0$, etc.).
Insofar as Eq.~\eqref{EQ:shape-parameter} is a fast-converging power-series in $x^\ddagger$, 
with each subsequent term much smaller than the previous, it makes sense
to view the subleading term on the right-hand-side of Eq.~\eqref{EQ:shape-parameter} 
as a proxy for those deviations:
\begin{equation} \label{EQ:shape-deviation}
\begin{split}
\frac{R^\ddagger_{1,3} x^\ddagger}{3\kappa^\ddagger} 
&= \frac{(R_l\kappa_b^3 + R_b\kappa_l^3)(x_b - x_l)}{3\kappa_l\kappa_b(\kappa_l+\kappa_b)^2}\\
&= 1 - \nu + \text{small corrections.}
\end{split}
\end{equation}
Hence, via Eq.~\eqref{EQ:extinction-force}, the extinction force can be approximated by
its quadratic-model value [Eq.~\eqref{EQ:extinction-force-quadratic}] up to rescaling
by a shape-dependent factor:
\begin{equation}
\begin{split}
\frac{\partial \Delta G^\ddagger}{\partial x^\ddagger} &= 
\kappa^\ddagger x^\ddagger
- \kappa^\ddagger x^\ddagger\frac{R^\ddagger_{1,3}x^\ddagger}{2\kappa^\ddagger} + \cdots \\
&=\kappa^\ddagger x^\ddagger \Bigl[ 1 - \frac{3}{2}(1-\nu) \Bigr] + \cdots
\approx \kappa^\ddagger x^\ddagger\biggl(\frac{3\nu-1}{2}\biggr).
\end{split}
\end{equation}
Typical values for smooth energy profiles ($2/3 \lesssim \nu \lesssim 6/5$)
suggest $0.5 \lesssim (3\nu-1)/2 \lesssim 1.3$, so we expect the
true extinction force value to be never more than a factor of two
away from $\kappa^\ddagger x^\ddagger$. Of course, when $U(x)$ is
known, it is straightforward to compute the exact extinction force numerically.

\subsection{Universality of the biased escape rate}

The calculations in this section are meant merely as a motivation
for the two-force-scale arguments we make in the paper. We 
assume Langevin behaviour with moderate to strong friction and
ignore the complications of non-Markovian dynamics.
Following Kramers, the escape rate from the left well of the untilted energy landscape is
\begin{equation} \label{EQ:Kramers-rate}
\begin{split}
k_0 &\propto \frac{1}{\sqrt{\kappa_l \kappa_b}} \exp\bigl(-\beta \Delta G^\ddagger\bigr)\\
&= \frac{1}{\sqrt{-U''(x_l)U''(x_b)}} \exp\bigl(-\beta[U(x_b)-U(x_l)]\bigr).
\end{split}
\end{equation}
The corresponding expression for the tilted case can be produced by
substituting $U(x_\alpha) \to \tilde{U}(\tilde{x}_\alpha)$.
If we expand around the $F=0$ case and collect terms order by order within the 
argument of the exponential, we arrive at
\begin{equation} \label{EQ:biased-escape-rate}
\begin{split}
k(F) = k_0 \exp
\biggl[
F&\biggl(\beta x^\ddagger + \frac{R^\ddagger_{1,2}}{2{\kappa^\ddagger}^2}
\biggr)\\
-F^2&\biggl(\frac{\beta}{2\kappa^\ddagger}
-\frac{R^\ddagger_{2,4}}{2{\kappa^\ddagger}^4}
+\frac{Q^\ddagger_{1,3}}{4{\kappa^\ddagger}^3}
\biggr)~~~~~~~~\\
-F^3&\biggl(\frac{\beta R^\ddagger_{1,3}}{6{\kappa^\ddagger}^3}
- \frac{2R^\ddagger_{3,6}}{3{\kappa^\ddagger}^6}
+ \cdots
\biggr)
+ O(F^4)
\biggr].
\end{split}
\end{equation}
In this equation, the terms proportional to $\beta$ come from the 
exponential in Eq.~\eqref{EQ:Kramers-rate}, whereas the 
temperature-independent contributions originate under the radical of the prefactor in Eq.~\eqref{EQ:Kramers-rate}. Since
\begin{equation}
\Delta G^\ddagger(F) = \Delta G^\ddagger(0) - Fx^\ddagger + \frac{1}{2\kappa^\ddagger}F^2
+ \frac{R_{1,3}^\ddagger}{6{\kappa^\ddagger}^3} F^3 + \cdots,
\end{equation}
Eq.~\eqref{EQ:biased-escape-rate} can also be expressed as
\begin{equation}
k(F) = k_0\exp(\beta\bigl[\Delta G^\ddagger(0) - \Delta G^\ddagger(F)
+ O(1/\beta,F)\bigr]).
\end{equation}
Although the terms replaced here by $O(1/\beta,F)$ may be small---in
the limit of very large barrier height ($\beta\Delta G^\ddagger(0) \gg 1$)
or possibly even for particular shapes of the energy landscape---they are
generally not negligible. As we will see, a renormalization of the
force scales $F_T$ and $F_B$ from their ``bare'' values is necessary
to absorb the discrepancy.

An important insight is that the logarithmic relative rate can be written in
the form
\begin{equation} \label{EQ:log-relative-rate}
Y(F) = \ln \frac{k(F)}{k_0} = \frac{F}{F_T} - \frac{F^2}{2F_TF_B} - \frac{CF^3}{2F_TF_B^2} + \cdots
\end{equation}
In this expression we have introduced two new dimensionful coefficients (with units of force),
defined according to
\begin{equation} \label{EQ:FT-FT-implicit-definitions}
\begin{split}
\frac{1}{F_T} &= \beta x^\ddagger 
+ \frac{R^\ddagger_{1,2}}{2{\kappa^\dagger}^2}\\
&= \beta x^\ddagger \biggl\{ 1 + \underbrace{\frac{R^\ddagger_{1,2}}{2\beta {\kappa^\dagger}^2 x^\ddagger}}_{\lambda_T}\biggr\}
\equiv \frac{1+\lambda_T}{F_T^{(0)}},\\
\frac{1}{F_TF_B} &= 
\frac{\beta}{\kappa^\ddagger}
-\frac{R^\ddagger_{2,4}}{{\kappa^\ddagger}^4}
+\frac{Q^\ddagger_{1,3}}{2{\kappa^\ddagger}^3}\\
&= \frac{\beta x^\ddagger}{\kappa^\ddagger x^\ddagger} \biggl\{ 1 - \underbrace{\frac{1}{\beta}\biggl(
\frac{R^\ddagger_{2,4}}{{\kappa^\ddagger}^3}
-\frac{Q^\ddagger_{1,3}}{2{\kappa^\ddagger}^2}}_{\lambda_{TB}}\biggr)\biggr\}
\equiv \frac{1-\lambda_{TB}}{F_T^{(0)}F_B^{(0)}},
\end{split}
\end{equation}
along with a dimensionless constant $C$. Matching the $O(F^3)$ terms in Eqs.~\eqref{EQ:biased-escape-rate}
and \eqref{EQ:log-relative-rate} and invoking Eq.~\eqref{EQ:shape-deviation}, we identify $C = 1-\nu + \cdots$, with 
the elision hiding additional terms that are shape and temperature dependent but small; specifically,
\begin{equation} \label{EQ:order-three-prefactor}
C = \frac{R_{1,3}^\ddagger x^\ddagger}{3\kappa^\ddagger} \frac{(1+\lambda_T)}{(1-\lambda_{TB})^2}
\biggl(1 - \frac{4R_{3,6}^\dagger}{3\beta R_{1,3}^\ddagger {\kappa^\ddagger}^3} + \cdots \biggr).
\end{equation}

The advantage of the rewriting in Eq.~\eqref{EQ:log-relative-rate} is that we have picked out two force scales,
$F_T$ and $F_B$, whose magnitude is determined---up to modest renormalization by $\lambda_T$
and $\lambda_{TB}$---by $F_T^{(0)} = 1/\beta x^\ddagger$ and $F_B^{(0)} = \kappa^\ddagger x^\ddagger$. 
Equation~\eqref{EQ:FT-FT-implicit-definitions} implies
\begin{equation}
F_T = F_T^{(0)}\biggl(\frac{1}{1+\lambda_T}\biggr),~~
F_B = F_B^{(0)}\biggl(\frac{1+\lambda_T}{1-\lambda_{TB}}\biggr).
\end{equation}
To give a more physical picture, we interpret $F_T$ as the typical thermal force that provides the kick out of the well
and $F_B$ as the applied force required to fully extinguish the barrier:
The ratio of the two force scales is
\begin{equation} \label{EQ:force-scale-ratio}
\frac{F_B}{F_T} 
= \beta\kappa^\ddagger {x^\ddagger}^2 \frac{(1+\lambda_T)^2}{(1-\lambda_{TB})}
= \frac{2\beta \Delta G^\ddagger}{\nu}(1 + \cdots).
\end{equation}

A key observation is that if we view the escape rate
as a function of a reduced applied force $f=F/F_B$, measured in units of the barrier extinction force scale, then
Eq.~\eqref{EQ:log-relative-rate} transforms to
\begin{equation} \label{EQ:log-relative-rate-expansion}
\begin{split}
Y(F)
\xrightarrow{F \to F_Bf} &\frac{F_Bf}{F_T} - \frac{(F_Bf)^2}{2F_TF_B}
- \frac{C(F_Bf)^3}{2F_TF_B^2} + \cdots\\
&\hspace{-1em}= \frac{F_B}{F_T}\biggl( f - \frac{1}{2}f^2 - \frac{1}{2}Cf^3 + \cdots \biggr).
\end{split}
\end{equation}
Note that the terms at order $f$ and $f^2$ are wholly independent 
of the details of the system. $(C/2)f^3$ is the leading nonuniversal term,
but even there [as per Eq.~\eqref{EQ:order-three-prefactor}] the shape dependence 
is quite weak and the temperature dependence almost negligible. 
This means that truly idiosyncratic contributions do not show up until order $f^4$,
and those we expect to be heavily suppressed just by power reduction; in practice, $f = F/F_B < 1$,
since escape almost always precedes complete elimination of the barrier.

Moreover, since physical considerations demand that the escape rate increase with $F$,
it is legitimate to apply a series acceleration transformation by which Eq.~\eqref{EQ:log-relative-rate-expansion}
is expanded in terms of some function of $f$ that is monotonic increasing but slower growing than the monomial $f$ itself;
one might consider $f/(1+f/2)$ (as in Ref.~\onlinecite{Adhikari-PRR-20}) or $\ln(1+f)$, say.
Then $f - (1/2)f^2 - (C/2)f^3 + \cdots$
can be recast as
\begin{equation} \label{EQ:universal-function-geometric}
\frac{f}{1+f/2} - \biggl(\frac{C}{2}+\frac{1}{4}\biggr)\biggl(\frac{f}{1+f/2}\biggr)^3  + \cdots
\end{equation}
or
\begin{equation} \label{EQ:universal-function-logarithm}
\ln(1+f) - \biggl(\frac{C}{2}+\frac{1}{3}\biggr)\bigl[\ln(1+f) \bigr]^3 + \cdots.
\end{equation}

Since we expect $-1/5 \lesssim C \lesssim 1/3$ (and often $\lvert C \rvert \ll 1$),
Eqs.~\eqref{EQ:universal-function-logarithm} and \eqref{EQ:universal-function-geometric}
are close to being universal even up to order three.
This leads us to posit that the logarithmic relative escape rate
has a form reminiscent of the finite-size scaling ansatz of
a critical state: viz., the form given by Eq.~\eqref{EQ:scaling-form}.
Then, since $F_T/F_B \approx 1/(\beta\kappa^\ddagger {x^\ddagger}^2) \approx \nu/(2\beta \Delta G^\ddagger) \ll 1$,
the quantity $(F_T/F_B)Y(F)$
should collapse onto a universal curve
when plotted against $f = F/F_B$:
\begin{equation} \label{EQ:data-collapse-without-subleading-corrections}
\begin{split}
\frac{F_T}{F_B}Y(F) = \frac{F_T}{F_B}\ln \frac{k(F)}{k_0} 
\approx \mathbb{Y}(f, 0).
\end{split}
\end{equation}
Other combinations of $(F_T/F_B)Y(F)$ may bring about an even cleaner coincidence.
For example,
\begin{multline} \label{EQ:data-collapse-with-subleading-corrections}
\ln(1+f)
= \frac{F_T}{F_B}Y(F) + \biggl(\frac{C}{2}+\frac{1}{3}\biggr)\bigl[\ln(1+f) \bigr]^3 + \cdots\\
=\frac{F_T}{F_B}Y(F)\biggl\{1 + \biggl(\frac{C}{2}+\frac{1}{3}\biggr)\biggl[\frac{F_T}{F_B}Y(F)\biggr]^2\biggr\} + \cdots.
\end{multline}

As for the utility, imagine that there is a set of escape 
rate measurements for which the underlying profile $U(x)$ is 
unknown. Then, even without a model or fitting form for $k(F)$,
we can still engineer graphical collapse of the data onto a
common curve by rescaling and careful adjustment of the free
parameters $F_T$ and $F_B$.

\subsection{Data collapse of the rupture force}

A population $n(t)$ of systems prepared in 
a well and subject to an escape rate $k(F)$
is subject to the rate equation $\dot{n} = -kn$.
If the pulling force increases linearly in time,
with a constant loading rate $KV$, then
\begin{equation} \label{EQ:population-decay}
\frac{dn}{dt} = -k(F)n(t) = -k(KVt)n(t).
\end{equation}

The time $\hat{t}$ for half the population to escape is
given by
\begin{equation} \label{EQ:half-life-defn}
\ln 2 = \int_1^{1/2} \frac{dn}{n} = - \int_0^{\hat{t}}\!dt\,k(KVt).
\end{equation}
In the Bell-Evans picture~\cite{Bell-Science-78,Evans-BPJ-91}, which supposes 
that the biased rate is simply $k(F) = k_0\exp (\beta F x^\ddagger)$,
Eq.~\eqref{EQ:half-life-defn} becomes
\begin{equation}
\ln 2 = \frac{k_0}{\beta KV x^\ddagger}\bigl( e^{\beta KV \hat{t} x^\ddagger} - 1\bigr)
= \frac{F_T^{(0)}k_0}{KV}\bigl( e^{\beta \hat{F} x^\ddagger} - 1\bigr).
\end{equation}
Then $\hat{F}$, the force at half-life, is
\begin{equation} \label{EQ:Bell-Evans-rupture-force}
\hat{F} = F_T^{(0)} \ln\biggl(1 + \frac{KV\ln 2}{F_T^{(0)}k_0} \biggr).
\end{equation}

On the other hand, if the escape rate is represented using the universal part of Eq.~\eqref{EQ:universal-function-logarithm},
via $\ln[k(F)/k_0]
=(F_B/F_T)(f - f^2/2 + \cdots)
=(F_B/F_T)\ln(1+f)$, then
\begin{equation} \label{EQ:rate-cf-Bell-Evans}
\begin{split}
k(F) &= k_0\exp\biggl[\frac{F_B}{F_T}\ln(1+f)\biggr] = k_0\biggl(1+\frac{F}{F_B}\biggr)^{\frac{F_B}{F_T}}\\
&= k_0\biggl(1+\frac{F}{\kappa^\ddagger x^\ddagger}\biggr)^{\beta \kappa^\ddagger {x^\ddagger}^2 + \cdots}\\
&= k_0\biggl[\biggl(1+\frac{F}{\kappa^\ddagger x^\ddagger}\biggr)^{\kappa^\ddagger x^\ddagger}\biggr]^{\beta x^\ddagger + \cdots}.
\end{split}
\end{equation}
The omitted terms [denoted by $\cdots$ in the exponent of the last line of Eq.~\eqref{EQ:rate-cf-Bell-Evans}] 
are ones that become negligible at low temperature and large barrier height;
in that same limit, we can formally take $\kappa^\ddagger x^\ddagger \to \infty$, which allows us
to recover the Bell-Evans expression, $k(F) \to k_0\exp(\beta F x^\ddagger)$.

We need not resort to such a limit, however, since the half-life can be solved analytically:
\begin{equation}
\begin{split}
\ln 2 &= \int_0^{\hat{t}}\!dt\,k(F(t)) = \int_0^{\hat{t}}\!dt\,k(KVt)\\
&= \int_0^{\hat{t}}\!dt\,k_0\biggl(1+\frac{KVt}{F_B}\biggr)^{\frac{F_B}{F_T}}\\
&= k_0F_B\biggl(\frac{ -1 + (1 + KV \hat{t}/F_B)^{1+F_B/F_T}}
{(1+F_B/F_T)KV}\biggr).
\end{split}
\end{equation}
This corresponds to an average rupture force
\begin{equation}
\begin{split}
\hat{F} &= KV\hat{t} = F_B\biggl\{ \biggl[1+ \biggl(1+ \frac{F_B}{F_T}\biggr)
\frac{KV\ln 2}{k_0 F_B}\biggr]^\frac{F_T}{F_T+F_B} - 1\biggr\}\\
&= \frac{(\ln 2)KV}{k_0} - \frac{(\ln 2)^2(KV)^2}{2F_Tk_0^2}\\
&\qquad\qquad + \frac{(2F_B + F_T)(\ln 2)^3(KV)^3}{6F_BF_T^2k_0^3} + \cdots.
\end{split}
\end{equation}
A useful resummation is
\begin{multline} \label{EQ:half-life-force-resummed}
\hat{F} = F_T\ln\biggl[1 + \frac{KV\ln 2}{F_Tk_0} \biggr]\\
+ \frac{2F_B-F_T}{6}\biggl(\ln\biggl[1 + \frac{KV\ln 2}{F_Tk_0} \biggr]\biggr)^3 + \cdots,
\end{multline}
the first term of which is identical to the right-hand-side of
Eq.~\eqref{EQ:Bell-Evans-rupture-force}, up to the renormalization $F_T^{(0)} \to F_T$.

In order to put Eq.~\eqref{EQ:half-life-force-resummed} into a scale-invariant
form, we define the half-life pulling force with respect to the thermal force
scale, $\hat{f}_T = \hat{F}/F_T$, and a dimensionless loading rate, $r_T = KV/(F_Tk_0)$.
This leads to
\begin{equation}
\begin{split}
\hat{f}_T &= \ln\bigl(1 + r_T\ln 2\bigr)\\
&\qquad + \frac{(2F_B/F_T-1)}{6}\Bigl[\ln\bigl(1 + r_T\ln 2\bigr)\Bigr]^3\\
&\approx \ln\bigl(1 + r_T\ln 2\bigr)
+ \frac{(2F_B/F_T-1)}{6}\hat{f}_T^3.
\end{split}
\end{equation}\\[0.1cm]
In general, $F_B/F_T \sim 2\beta \Delta G^\ddagger/\nu$ is not small.
But so long as $(F_B/F_T)\hat{f}_T^2 = F_BF^2/F_T^3 \ll 1$, it is
appropriate to write
\begin{equation}
\hat{f}_T \biggl(1 - \frac{(2F_B/F_T-1)}{6}\hat{f}_T^2\biggr) = \ln\bigl(1 + r_T\ln 2\bigr).
\end{equation}

\section{Numerical results \label{SEC:numerical-results}}

We carried out a thorough and comprehensive Langevin simulation study.
At the start of each run, the system was prepared in a properly equilibrated
state: an initial position and velocity were drawn from the heat bath
distribution of the appropriate energy profile, with the constraint that the
particle be situated on the originating-well side of the barrier. Forward
evolution was carried out with adaptive time steps taken small enough that the
discretization error could be shown to be negligible. The simulation made use
of a high-quality, long-period pseudorandom number generator that guaranteed
the statistical independence of the instantaneous thermal forces
(based on Gaussian-distributed noise $\xi(t)$ that is unbiased, 
$\langle \xi(t) \rangle = 0$, and uncorrelated except at identical time 
slices, $\langle \xi(t) \xi(t') \rangle = \delta(t-t')$).

We considered seven different potentials having shape 
parameter~\cite{Dudko-PRL-06} $\nu = 0.66$, 0.75, 0.83, 0.9, 1, 1.1, 1.2; 
these values step through the full range of possibilities for smooth potentials 
based on polynomials. This family of energy potentials---translated and
rescaled to coincide at the bottom of the originating well and at the top of
the barrier so as to emphasize the shape difference---is displayed in 
the lower panel of Fig~\ref{FIG:Double-well}. We also considered eight barrier 
regimes, with $1/\beta \Delta G^\ddagger = k_BT/\Delta G^\ddagger$ taking 
values 0.25, 0.3, 0.35, 0.40, 0.45, 0.50, 0.55, 0.60, a list that includes 
temperatures high enough (or, equivalently, barriers low enough) to be outside 
the range of validity for pure Arrhenius-law behavior.  Simulations were 
carried out in both the constant-force and steady-loading modes, with relative 
applied forces ($F/F_B$) and relative loading rates ($KV/k_0F_T$) each spanning 
nearly ten orders of magnitude. For each run, the trajectory leading to barrier 
traversal was captured and analyzed.

\begin{figure*}
\begin{center}
\includegraphics{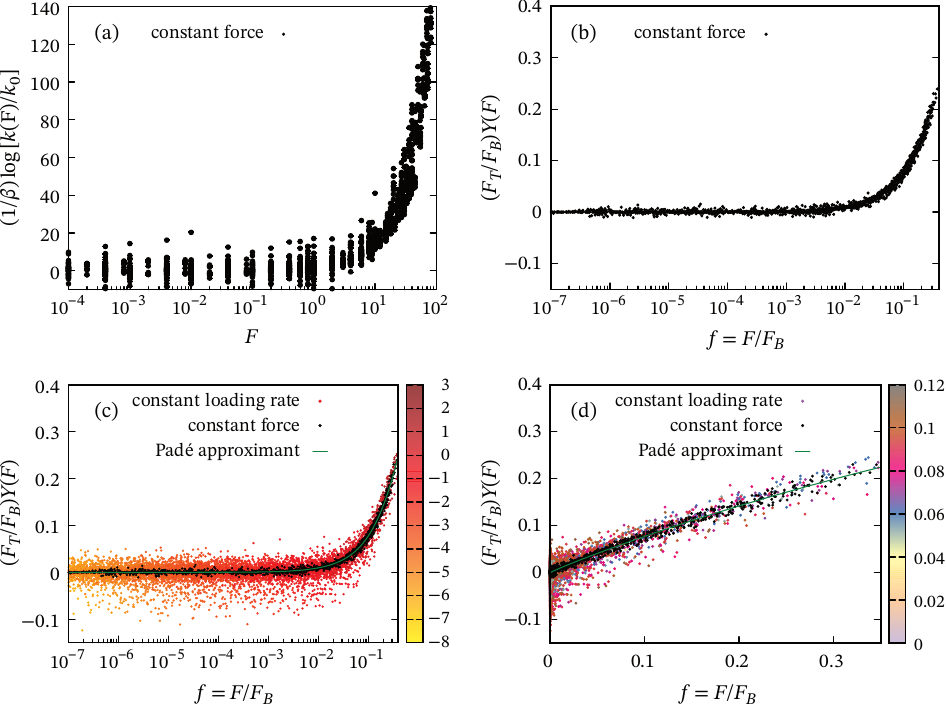}
\end{center}
\caption{\label{FIG:data-collapse-relative-rate}
The upper two panels show constant-force data only, (a) unscaled 
and raw and (b) rescaled to produce the predicted data collapse. The lower 
panels, like panel (b), 
show plots of $(F_T/F_B)Y(F) = (F_T/F_B)\ln{(k(F)/k_0)}$ versus 
$f = F/F_B$. These include all the output generated by our simulations 
(from both forcing protocols) and offer different views of the same underlying 
data set. Black circles correspond to simulations executed in the 
constant-force mode. Colored solid circles denote data from steady-loading 
runs. The green line is a low-order Pad\'e approximant fit through the 
data points. (c) The horizontal axis uses a logarithmic scale. Color 
saturation increases with the relative pulling rate, $r_T = KV/(k_0F_T)$. 
Numbers on the palette legend refer to the order of magnitude, $\ln r_T$. 
(d) The horizontal axis is linear. Colors now represent
$F_T/F_B \approx 1/\beta\kappa^\ddagger {x^\ddagger}^2
\approx \nu/2\beta\Delta G^\ddagger$, which characterizes the barrier regime.}
\end{figure*}

\begin{figure}
\begin{center}
\includegraphics{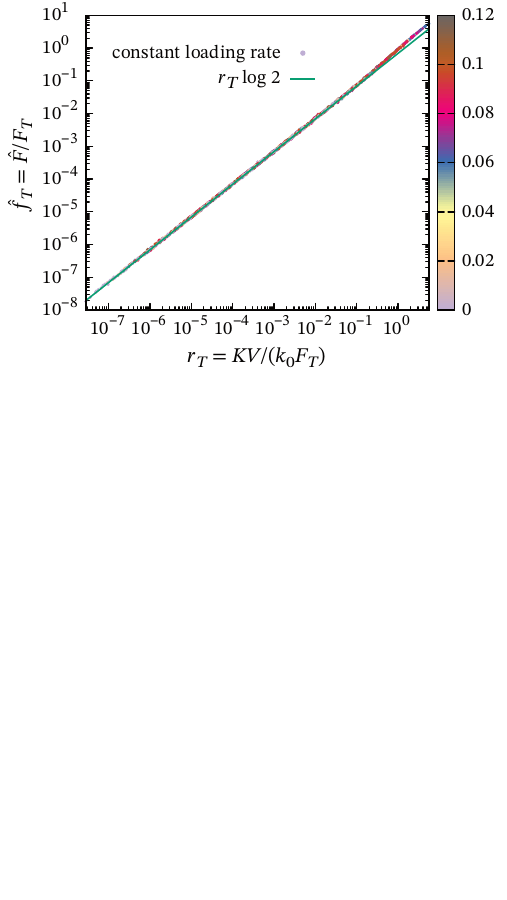}
\end{center}
\caption{\label{FIG:data-collapse-half-life} 
These data are from simulations performed with dynamic tilting of the potential at a constant loading rate.
Shown here is $\hat{f_T} = \hat{F}/F_T$, the force at half-life measured with respect to the thermal force scale,  
plotted against the effective pulling rate, $r_T = KV/(k_0F_T)$.
Each data point is a solid circle, colored as per the legend according to its $F_T/F_B$ value.}
\end{figure}

The numerical simulations were carried out using a modified version~\cite{Gronbech-Jensen-Mol-13} of the standard 
Verlet algorithm~\cite{Verlet-PRL-67}. In each run, a critical time $t_c$, taken to be either the first-passage time of the
particle over the barrier or the moment at which the barrier vanished, was recorded. For each energy potential profile and 
simulation mode, 3000 instantiations were generated.

In the constant-force mode, the rate $k(F)$ was computed from the mean escape time: $k(F) =1/t_{\text{avg}}$, 
where $t_{\text{avg}} = (1/3000)\sum_{i=1}^{3000}t_c^{(i)}$.
In the steady-loading mode, the linear correspondence $F_c = KVt_c$ gave rise to 3000 critical force values, on the basis
of which further analyses were performed. First, the $F_c$ values were sorted to identify their median value, which corresponds
to the half-life force $\hat{F}$ (the force at which half of a population of independent particles would have escaped the well). 
Second, the $F_c$ values were bootstrapped~\cite{Efron-CRC-93} to obtain 
the cumulative probability distribution $P(F_c) = \int_0^{F_c}\!dF\,p(F)$ and probability density $p(F_c) = P'(F_c)$. Finally the value of $k(F)=k(KVt)$, the instantaneous rate of barrier crossing at a particular bias strength, was obtained using the relation $k(F_c) = {KVp(F_c)}/{(1-P(F_c))}$~\cite{Dudko-PRL-06}.

The next step was to test the universality proposition by graphical means. We found strong evidence
in its favor: the data collapse predicted by Eq.~\eqref{EQ:scaling-form}
is revealed in Fig.~\ref{FIG:data-collapse-relative-rate}.
In order to perform the conversion to reduced variables, each data point was associated with an individualized value of
$F_B$ and $F_T$. The former was obtained numerically, simply by solving for the applied
force required to extinguish the barrier;
the latter was estimated according to
\begin{equation} \label{EQ:thermal-force}
\frac{1}{F_T} = \beta x^\ddagger 
+ \frac{R^\ddagger_{1,2}}{2{\kappa^\dagger}^2}
= \beta (x_b-x_l)
+ \frac{1}{2}\biggl(\frac{R_b}{\kappa_b^2} + \frac{R_l}{\kappa_l^2}\biggr).
\end{equation}
Here, we have made use the notation introduced in 
Sec.~\ref{SUBSECT:higher-order-corrections}.

Figure~\ref{FIG:data-collapse-relative-rate}(a) presents a linear-log plot of $(F_T/F_B)Y(F) = (F_T/F_B)\ln{(k(F)/k_0)}$ versus $ f = F/F_B$. The data points for the steady-loading analysis are colored
according to the simulation-specific loading rate, and one can observe the smooth progression of data-point placement,
weak loading to strong, tracing out the universal curve from left to right. The constant-force data (black circles)
show considerably less scatter, but the two data sets are remarkably consistent.
What makes this result so compelling is that the data collapse holds over a huge diversity of energy profiles
and simulation conditions. We also remark that the steady-loading and constant-force approaches require quite
different styles of simulation and analysis, but both yield the same underlying curve;
Pad\'{e} approximants fit to one or the other data set produce nearly identical functions.

Figure~\ref{FIG:data-collapse-relative-rate}(b) shows the same data plotted on a linear scale. 
This view highlights the behavior at large forces, a regime in which the barrier is already substantially
reduced at the time of barrier traversal. Here, the false color emphasizes the diversity
in barrier height regimes, and we can see that data collapse holds over a wide range of ratios $F_T/F_B$.

Figure~\ref{FIG:data-collapse-half-life} presents a wholly different data collapse scheme,
based only on simulations performed in the steady-loading mode.
There, the reduced half-life force $\hat{f_T }= \hat{F}/F_T$ is plotted versus the reduced loading rate $r_T = KV/(k_0F_T)$. 
It is worth emphasizing again that the complete data set comes from simulations with seven different potential landscapes 
covering the full range of plausible $\nu$ values, $1/(\beta \Delta G^\ddagger)$ ranging from $0.2$ to $0.6$, 
and loading rates running from $r_T = 10^{-8}$ to $100$. Despite encompassing a large collection of different
systems in distinct physical regimes, these data show an astonishing degree of collapse.

\section{Conclusion \label{SEC:conclusion}}

We have argued for universality in the biased activated-barrier-crossing 
problem and presented strong numerical evidence in favor of the existence 
of some underlying scaling function for $Y(F) = \ln[k(F)/k_0]$. Our 
simulated data show collapse onto a single curve when recast into suitably
reduced coordinates. This is true for data generated in simulations operating 
over a wide range of bath temperatures, applied forces, and loading rates 
and over a family of potential landscapes with different underlying barrier
shapes.

These observations suggest the utility of data collapse as a practical tool
for analysis. While the original motivation for this work was the mechanical
unfolding of biopolymers, the universality we have identified is widely
relevant. It applies to situations across many branches of science, wherever
the energy landscape picture is germane and the experimental setup involves
barrier traversal assisted by active pulling. Our recommendation is that
measurements of well-escape statistics be transformed to identify best
values of the intrinsic force scales (from which can be inferred some 
combination of $x^\ddagger$, $\kappa^\ddagger$, $\Delta G^\ddagger$, and 
$\nu$). $F_T$ and $F_B$ are to be treated as free parameters and tuned until 
data collapse is achieved and the universal curve emerges.

To motivate our general approach, and to provide logical and formal
scaffolding for the scaling argument, we have relied on analytic expressions for 
the biased escape rates. In Sec.~\ref{SEC:theoretical-motivation}, these were 
computed within the context of Kramer's theory---in  particular, the most 
straightforward version, which presumes moderate, Ohmic friction, does not include 
explicit finite-barrier corrections~\cite{Pollack-CPC-14}, and does not attempt a 
more sophisticated reconsideration of the Kramers-Grote-Hynes transmission
factor~\cite{Pollak-CPC-23}. 

Our demonstration of scaling, however, does not depend on Kramer's theory specifically
nor on the quality of the analytical approximation. We emphasize that the data collapse 
in this work is achieved with escape rates determined {\it empirically} from Langevin simulations 
(as described in Sec.~\ref{SEC:numerical-results}). That is to say, the scaling viewpoint
we advocate is agnostic with respect to the well-escape model. We simply point
out that, in principle, it should be possible to plot experimental measurements of the 
escape rate in reduced coordinates such that the data falls on a single curve; this
relies on identifying system-specific values of $F_T$ and $F_B$,
the two force scales whose values must be varied to produce the data collapse.
Details of the underlying potential can then be inferred from $F_T$ and $F_B$, although that
final step does introduce some dependence on how escape from the well is modeled.

We close with one final comment. It has already been established that 
the barrier vanishes generically according to 
$\Delta G^\ddagger \sim (F_B - F)^{3/2}$ for bias forces in close 
vicinity of the barrier extinction force 
$F_B$~\cite{Maloney-PRE-06,Lacks-JPCB-10}. This is an important result.
It leads to predictable, standardized behavior in that force regime
and thus provides the basis for a kind of force 
spectroscopy~\cite{Dudko-PNAS-03}. Nonetheless, we view this as 
{\it asymptotic} behavior (as $F \to F_B$) and not an example of 
{\it universality} in the traditional sense.
The results reported in this paper are applicable to 
all applied forces less than $F_B$ and do not dependent on any 
particular force limit. The data collapse, for instance, appears
to be valid for $f = F/F_B$ ranging at least from $10^{-7}$ to $0.4$.
The universality we advocate for here is an underlying commonality 
across many orders of magnitude that is revealed by rescaling.

\end{document}